\documentclass[aps,showpacs,amsmath,amssymb,12pt]{revtex4}
\usepackage{latexsym}
\usepackage{bm}

\def\HollowBox #1#2{{\dimen0=#1 \advance\dimen0 by -#2
       \dimen1=#1 \advance\dimen1 by #2
        \vrule height #1 depth #2 width #2
        \vrule height 0pt depth #2 width #1
        \llap{\vrule height #1 depth -\dimen0 width \dimen1} 
       \hskip -#2
       \vrule height #1 depth #2 width #2}}
\def\BOX{\HollowBox{.100in}{.010in}}

\begin{document}

\title{Chern-Simons Modified General Relativity: Conserved charges }

\author{Bayram Tekin}  
\email{btekin@metu.edu.tr}
\affiliation{Department of Physics, Middle East Technical University, \\
 06531, Ankara, Turkey}

\date{\today}

\begin{abstract}
We construct the conserved charges ( mass and angular momentum)
of the Chern-Simons modified General  Relativity  in asymptotically 
flat and Anti-de Sitter (AdS) spacetimes. Our definition is based on 
background Killing symmetries and reduces to the known expressions in the 
proper limits.

\end{abstract}


\maketitle

\section{\label{intro} Introduction}

Soon after General Relativity (GR) was written, various 
modifications of it- such as adding higher curvature terms based on the 
invariants of the Riemann tensor- or adding scalar fields appeared. In 
the beginning, these attempts usually arose from pure academic interest 
and 
eventually tried to answer the question of how unique GR was. The simplest 
modification of GR, adding a cosmological constant to the action, was done by 
Einstein himself ( although his reason of bringing in a cosmological constant 
was more physical than of academic curiosity.)  
More recently, modifying gravity at various energy scales has become, one 
can safely say, a necessity. In the solar system, for which we have plenty 
of 
experimental tests, GR works perfectly well and so there is no compelling 
reason to modify it. But at very short and large distances, we know that 
GR cannot be the whole story and hence follows the various modifications and generalizations.

In the literature, most of the modifications of GR do not change the 
symmetries, such as parity invariance {\it {etc.}}. In this paper, we 
shall be interested in a more recent, rather 
unconventional yet very intriguing modification introduced by 
Jackiw and Pi \cite{jp}. These authors define a new {\it{four}} dimensional, 
symmetric tensor which is analogous to the {\it{three}} dimensional 
Cotton tensor ( or Chern-Simons term in the action) which gave rise to the celebrated 
Deser-Jackiw-Templeton`s theory 
of Topologically Massive Gravity (TMG) \cite{djt}. TMG and its abelian and 
non-abelian cousins ( TM electrodynamics and 
TM Yang-Mills 
theory )have been much studied: But, since these are three 
dimensional theories, they do not shed much light on the problems of 
the four 
dimensional world, they are more relevant to the planar condensed 
matter systems. On the other hand, upgrading the Chern-Simons term to 
four dimensions could actually yield interesting physics; such as Lorentz 
violating gravity and so on. In fact, in the context of Maxwell`s theory, 
such a modification was done some time ago by Carroll {\it{et. al}} \cite{cfj} 
who obtained a Chern-Simons modified electrodynamics which apparently was 
falsified by measurements of light from distant galaxies \cite{cfj}.

In any matter-coupled gravity theory, that is based on Riemannian 
geometry, field equations relate at {\it{each point}} in space a covariantly 
conserved symmetric tensor, obtained from 
the metric tensor and its derivatives, to the local energy momentum tensor 
$\tau_{\mu \nu}(x)$ of matter fields. Since in a diffeomorphism invariant 
theory, the coordinates $x^\mu$ could be traded with some new ones, one 
cannot 
define a meaningful ( that is gauge-invariant) local energy expression. But 
for certain spacetimes one can define total mass and angular momenta 
such as the Arnowitt-Deser-Misner ( ADM) \cite{adm} mass in asymptotically
 flat spaces or the Abbott-Deser (AD) \cite{ad} mass for a
asymptotically Anti-de Sitter (AdS) 
spaces. Mass and angular momentum in this construction is given in terms of 
asymptotic geometry and background Killing vectors. This approach has 
been quite useful and was carried out in detail for various higher 
derivative gravity models 
beyond 
(cosmological ) general relativity: For example conserved charges 
( in any coordinates !) of  higher curvature models were defined in 
\cite{dt1} and TMG`s charges were defined in \cite{dt2}. These 
definitions were employed \cite{dkt} to correctly compute the mass and angular 
momenta of the recently found $D$-dimensional Kerr-AdS black holes 
\cite{gibbons} and the supersymmetric solution of the TMG \cite{ost}. 
For a rigorous definition of asymptotically AdS spaces and a canonical 
construction of charges, we refer the reader to  
\cite{hete}. In this 
paper, we will construct the conserved charges of the Chern-Simons (CS) 
modified 
GR along the lines developed in the above mentioned works. As we shall see, 
it is a somewhat non-trivial problem and a brute-force approach of getting the 
charges in generic coordinates for both asymptotically flat and 
asymptotically AdS spaces is most certainly bound to fail unless some 
simplification  techniques that proved very handy in \cite{dt1,dt2} 
are employed. 

The outline of the paper is as follows: we first review the basics of the CS 
modified GR and recall some recently found facts about this model and 
then we briefly summarize how conserved charges, arising from background 
Killing vectors, can be constructed in generic gravity theories. 
Finally we find a general formula for Killing charges in the CS modified 
(cosmological) GR.  Our formula is generic enough to give the mass of the 
asymptotically flat and asymptotically AdS spacetimes as for the angular 
momentum, we only provide the formula for asymptotically flat backgrounds.

\section{\label{review} CS modified GR }

Here we introduce and quote the very basics of the model and refer the 
reader to the original paper \cite{jp} and  a nice recent review
 \cite{jackiw} or 
to \cite{gh}. [This latter work has constructed a Papapetrou type 
pseudo energy-momentum {\it{tensor}} which is {\it{not}} what we are
about to construct here. What we will find is a background conserved, 
coordinate independent total mass (as a component of a four vector, not a 
tensor ) and the angular momentum in asymptotically 
AdS or asymptotically flat spaces.] On the left-hand side of the field 
equations, we have the usual Einstein 
tensor plus a new symmetric tensor:     
\begin{equation}
G_{\mu \nu } + C_{\mu \nu} = 8 \pi G \tau_{\mu \nu}\, ,
\label{eincot}
\end{equation}
where the ``Cotton'' term reads
\begin{equation} 
C^{\mu \nu} = -\frac{1}{2 \sqrt{-g}} \Big( \,v_\sigma( \epsilon^{\sigma 
\mu 
\alpha \beta}\nabla_\alpha R^\nu\,_\beta + \epsilon^{\sigma \nu 
\alpha \beta}\nabla_\alpha R^\mu\,_\beta   ) +\nabla_\sigma v_\tau \, 
(\ast R^{\tau \mu \sigma \nu} +\ast R^{\tau \nu \sigma \mu}) \Big) \, .
\label{cotton}
\end{equation}
The dual Riemann tensor is defined as
\begin{equation}
\ast R^{\tau}\,_\sigma\,^{\mu \nu} \equiv \frac{1}{2}\epsilon^{\mu \nu \alpha 
\beta }\, R^\tau\,_{\sigma \alpha \beta},
\label{dualrieman}
\end{equation}
and the four-vector $v_\sigma$, coined as the {\it{embedding coordinate}}, 
is taken to be timelike and {\it \it{constant}}: $v_\sigma = 
(\frac{1}{\mu},0,0,0)$ to carry out the analogy with the well-known three 
dimensional TMG \cite{jp}. 
 
Of course, one would like to derive field equations from an action, which in 
this case is quite simple and well-known 
\begin{equation}
S =\frac{1}{16 \pi G} \int \, d^4x 
\left( \sqrt{-g} R + \frac{1}{4}\theta(x)\ast R R \right)\, .
\label{action}
\end{equation}
For constant $\theta$, the second term does not contribute to the field 
equations since it can be written as a boundary term. For $\theta(x) = 
v_\sigma x^\sigma$, minimization yields the field equations 
(\ref{eincot}). This is actually a somewhat 
unconventional theory since there appears a non-dynamical external `field`. 
[ See \cite{mariz1,mariz2} which discuss how a Lorentz-violating term such 
as above can be radiatively generated in quantum theory.]
Of course, we could consider adding  a kinetic term to $\theta(x)$ and 
not restrict it to the above form. 
Then,  we would have an axion field coupled to gravity which apparently 
can be obtained 
from string theory \cite{cdko}. ( See also the Appendix of \cite{seck} for 
a nice 
derivation of this.) For our purposes, we shall take $\theta(x)$ to be a 
non-dynamical external field, since getting conserved charges in the dynamical 
model is not too different from what we shall do below.

There are at least two important problems with (\ref{eincot}).  
First of all, unlike GR, these equations 
are incomplete in the sense that the covariant divergence of the added piece
 is non-zero and one has an extra constraint:
\begin{equation}
\nabla_\mu C^{\mu \nu}= \frac{1}{8 \sqrt{-g}} v^\nu \ast R R.
\label{constraint}
\end{equation}
So, for consistency solutions of this model should satisfy $\ast R R =0$ 
and therefore the usual Kerr metric does not solve 
the field 
equations but the Schwarzschild black hole  does \cite{jp}. Perturbation 
theory 
around the Schwarzschild solution ruled out small angular 
momentum for time-like $v_\sigma$ recently \cite{kmt}.[ Thus, one still 
needs to 
look for a Kerr type rotating solution for this theory.] We 
will see if and how the  
constraint (\ref{constraint}) will play a role in the construction 
of conserved charges. 
Second problem is that there is an explicit function $\theta(x)$ 
which seems to ruin our hope of finding conserved mass and angular momentum. 
Of course, for a generic external field, the theory will have no conserved 
charges.  But, if the function is chosen as above 
( namely, $\theta(x) =  t/ \mu$) then, both 
mass and angular momentum will be conserved, since  constant shifts in time ( 
$t \rightarrow t + t_0$) are still symmetries of the theory ( owing to the 
fact that $\ast R R$ is a boundary term)   
 and pure spatial rotational symmetries also survive since $\theta(x)$ was 
chosen to be time like and independent of $\vec{x}$.

As promised above, we now add a cosmological constant and consider 
\begin{equation}
G_{\mu \nu } + C_{\mu \nu}+ \Lambda g_{\mu \nu} = 8 \pi G \tau_{\mu \nu}\, 
.
\label{cosmocot}
\end{equation}
Note that even if we are interested in the $\Lambda =0$ theory, for the 
sake of computations, it is 
actually better to start with (\ref{cosmocot}) since then, one 
relatively easily ends up 
with conserved charges in generic coordinates as opposed to the Cartesian 
coordinates. Even before declaring what $v_\sigma$ is, it is easy to check 
that global AdS
\[ \bar{R}_{\mu\alpha\nu\beta} = \frac{\Lambda}{3} \, 
(\bar{g}_{\mu\nu} \, \bar{g}_{\alpha\beta} - 
\bar{g}_{\mu\beta} \, \bar{g}_{\alpha\nu}) \, , \quad
\bar{R}_{\mu\nu} = \Lambda \, \bar{g}_{\mu\nu}  \quad
 \, , \]
is a vacuum ( $\tau_{\mu \nu }=0$ ) solution since we have 
$\ast \bar{R}^{\tau}\,_\sigma\,^{\mu \nu} = \frac{\Lambda}{3}
\epsilon^{\mu \nu \tau}\,_ \sigma$. ( Note I will denote all the vacuum, or 
the background quantities with a bar.) 
Although, I will not prove here, it is straightforward yet somewhat 
lengthy to show that  AdS-Schwarzschild black hole is also a solution to 
our full theory.

\section{\label{kill} Killing Charges}

The general formalism of constructing ordinarily conserved charges ( as opposed 
to covariantly conserved ones)
based on background  Killing vectors was given in \cite{dt1}. 
Here we recapitulate some of 
the material which we shall use. We will be working with a $(-,+,+,+)$ metric
and our sign conventions are $[\nabla_\mu, \nabla_\nu]V_{\lambda} \equiv 
R_{\mu \nu \lambda}\,^\sigma V_\sigma $ and 
$R_{\mu \nu} =  R_{\mu \lambda \nu}\,^\lambda$. 
 
Let $\bar{g}_{\mu \nu}$ denote the background metric for the AdS 
( or flat space) that solves our full equations without $\tau_{\mu \nu}$ 
and let   
\begin{equation} 
g_{\mu \nu} \equiv \bar{g}_{\mu \nu} + h_{\mu \nu}\, ,
\end{equation}
solve the equations with a non-vanishing matter density. This equation is 
exact and $h_{\mu \nu}$ need not be small everywhere but we require that 
at infinity, away from the sources,  it goes to zero. ( Otherwise, one cannot 
consider $\bar{g}_{\mu \nu}$ as a background spacetime with zero charges and 
$h_{\mu \nu}$ as a perturbation outside a compact region, whose charge is 
measured with respect to the background.) 

Let $T_{\mu \nu }(h)$ denote the matter $\tau_{\mu \nu}$ plus all the 
higher power terms in $h$. Linearizing the full equations we have:
\begin{equation}
T_{\mu \nu }(h) = R^L\,_{\mu \nu} -\frac{1}{2}\bar g_{\mu \nu}R^L - 
\Lambda h_{\mu \nu} + C^L\,_{\mu \nu}\, .
\end{equation}
Some new notation will help us later, let us introduce
\begin{equation}
{\cal G}_{\mu\nu} \equiv  R_{\mu\nu} -\frac{1}{2}g_{\mu \nu} R+ 
\Lambda \, g_{\mu\nu}\, ,
\label{neweinstein}
\end{equation}
whose linear part reads as
\begin{equation}
{\cal G}_{\mu\nu}\,^L =  R^L\,_{\mu \nu} -\frac{1}{2}\bar g_{\mu \nu}R^L - 
\Lambda h_{\mu \nu}\, .
\label{linearneweinstein}
\end{equation}
As can be checked, this linearized tensor is background covariantly 
conserved: 
$\bar{\nabla}_\mu {\cal 
G}^{\mu\nu}\,_L =0 $. So, in order to now have $\bar{\nabla}_\mu 
T^{\mu\nu}\,_L =0$, we should have  $\bar{\nabla}_\mu
C^{\mu\nu}\,_L =0$. To see that this holds, we can linearize the 
consistency condition (\ref{constraint}) which upon explicit use of 
\begin{equation}
(R^\mu\,_{\alpha \beta \nu})_L = \bar{\nabla}_\beta 
\, \delta \Gamma^\mu\,_{\alpha \nu} -
\bar{\nabla}_\nu \, \delta \Gamma^\mu\,_{\alpha \beta},
\end{equation}
and
\begin{equation}
(\ast R^{\tau \mu \sigma \nu})_L = \frac{1}{2}\epsilon^{\sigma \nu \alpha 
\beta} \bar{g}^{\mu \rho} (R^\tau\,_{\rho \alpha \beta })_L -\frac{\Lambda}{3} 
\epsilon^{\sigma \nu \tau}\,_\rho \, h^{\rho \mu}\, ,
\end{equation}
gives 
\begin{equation}
\bar{\nabla}_\mu C^{\mu\nu}\,_L =0\, .
\end{equation}
Thus, at the linearized level, the consistency condition 
(\ref{constraint}) 
does {\it {not}} put any 
constraint on the metric. This fact is vital in constructing 
conserved charges. Let us explain how: if we have background symmetries as 
generated by 
background Killing vectors $\bar{\xi}^I_\mu$
\begin{equation}
\bar{\nabla}_\mu \xi^I_\nu + \bar{\nabla}_\nu \xi^I_\mu =0, 
\end{equation}
where $I$ refers to different Killing vectors  such as mass and angular 
momentum ( I shall suppress this 
index in what follows): we can find  partially conserved four-currents
\begin{equation}
\sqrt{-\bar{g}} \bar{\nabla}_\mu \Big( T^{\mu \nu}\, \bar{\xi}_\nu \Big) 
= \partial_\mu \Big(\sqrt{-\bar{g}} 
T^{\mu \nu}\, \bar{\xi}_\nu  \Big) = 0.
\end{equation} 
As usual, upon integration over the spatial 3-space  
and assuming that the three vector current falls of sufficiently fast at 
infinity, we will get conserved charges for the $\mu =0$ 
component, which up to a normalization reads
\begin{equation}
Q^0 = \int d^3\, x \sqrt{-\bar{g}} T^{0 \nu}\bar{\xi}_\nu\, .
\end{equation}
 Given background symmetries, 
and a divergence-free
energy-momentum tensor of matter fields, this approach guarantees the 
existence of the charges but getting them explicitly as surface integrals 
is a different issue which we shall work out below. [The reader might 
wonder why we have to convert the bulk integrals to surface integrals: In 
principle, we can of course work with the volume integrals but since in 
the bulk, there usually will be singularities of the spacetime, explicit 
computation of charges could become somewhat tricky. We can give the 
classical Maxwell theory as a simple analogy: volume integral of the 
divergence of the electric field has Dirac Delta functions coming from the 
points where the source charges are located. Hence we use the surface 
integral of the electric field ( the Gauss law) to count the charges. ]

We have to write $\sqrt{-\bar{g}}_\mu T^{\mu
\nu}\, \bar{\xi}_\nu $ as a surface integral at spatial infinity. 
The cosmological Einstein part was taken care of in \cite{dt1} or in a 
different notation earlier in \cite{ad}. Here we quote the final result 
\begin{eqnarray}
2\sqrt{\-\bar{g}}\,{\cal G}^{\mu \nu}_L \,\bar{\xi}_\nu
& = &\sqrt{\-\bar{g}}
\bar{\nabla}_\rho\left( \bar{\xi}_{\nu} \, \bar{\nabla}^{\mu} \, h^{\rho \nu} -
\bar{\xi}_{\nu} \, \bar{\nabla}^{\rho} \, h^{\mu\nu} +
\bar{\xi}^{\mu} \, \bar{\nabla}^{\rho} \, h -
\bar{\xi}^{\rho} \, \bar{\nabla}^{\mu} \, h \right.  \nonumber \\
& & \quad \qquad \left. + h^{\mu\nu} \, \bar{\nabla}^{\rho} \, 
\bar{\xi}_{\nu}
- h^{\rho \nu} \, \bar{\nabla}^{\mu} \, \bar{\xi}_{\nu}
+ \bar{\xi}^{\rho} \, \bar{\nabla}_{\nu} \, h^{\mu\nu}
- \bar{\xi}^{\mu} \, \bar{\nabla}_{\nu} \, h^{\rho \nu}
+ h \, \bar{\nabla}^{\mu} \, \bar{\xi}^{\rho}\right),
\label{einsteinboundary1}
\end{eqnarray}
where $ h = \bar{g}^{\mu \nu} h_{\mu \nu}$. To get this expression one 
makes 
use of the linearized Ricci tensor
\[ R_{\mu\nu}^{L} = \frac{1}{2} (- \bar{\BOX} \, {h}_{\mu\nu}
- \bar{\nabla}_{\mu} \, \bar{\nabla}_{\nu} \, h + \bar{\nabla}^{\sigma} \,
\bar{\nabla}_{\nu} \, h_{\sigma\mu} + \bar{\nabla}^{\sigma} \,
\bar{\nabla}_{\mu} \, h_{\sigma\nu}) \, , \]
and the linearized Ricci scalar
\[ R^{L} \equiv (R_{\mu\nu} \, g^{\mu\nu})^{L} =
R_{\mu\nu}^{L} \, \bar{g}^{\mu\nu} - \Lambda \, h = - \bar{\BOX} \, h
+ \bar{\nabla}_{\mu} \, \bar{\nabla}_{\nu} \, \bar{h}^{\mu\nu}
-  \Lambda \, h \, . \] 
Let us now concentrate on the new four-dimensional Cotton part. But first, 
we get a great deal of simplification if we re-write the Cotton term 
(\ref{cotton}) in the following form using (\ref{neweinstein})
\begin{equation}
C^{\mu \nu} = -\frac{1}{2 \sqrt{-g}}\Big( \,v_\sigma( \epsilon^{\sigma \mu
\alpha \beta}\nabla_\alpha {\cal G}^\nu\,_\beta + \epsilon^{\sigma \nu
\alpha \beta}\nabla_\alpha {\cal G}^\mu\,_\beta   ) +\nabla_\sigma v_\tau 
\, (\ast R^{\tau \mu \sigma \nu} +\ast R^{\tau \nu \sigma \mu}) \Big).
\label{cotton2}
\end{equation}
One should observe the nice cancellation in the $\frac{1}{2}g_{\mu \nu} R$ 
parts. 
The linearization of (\ref{cotton2}) about AdS yields
\begin{equation}
-2 \sqrt{-\bar{g}} C^{\mu \nu}_L = v_\sigma \epsilon^{\sigma \mu \alpha \beta} 
\bar{g}^{\nu \rho}\bar{\nabla}_\alpha {\cal G}^L\,_{\rho \beta} +v_\sigma \epsilon^{\sigma \nu \alpha \beta} 
\bar{g}^{\mu \rho}\bar{\nabla}_\alpha {\cal G}^L\,_{\rho \beta}
+\bar{\nabla}_\sigma v_\tau   
(\ast R^{\tau \mu \sigma \nu}\,_L +
\ast R^{\tau \nu \sigma \mu}\,_L ) \, .
\label{linearcotton}
\end{equation}
At this point, it is better to separate the last term involving the linearized 
dual Riemann tensors from the first two terms. One can immediately see that, 
for flat backgrounds in the Cartesian coordinates, the last term will not 
contribute to the charges, since $\bar{\nabla}_\sigma v_\tau = 
\partial_\sigma v_\tau =0$. For AdS backgrounds, on the other hand, the 
discussion bifurcates: the last term does not contribute to mass but it does 
contribute to the angular momentum. Thus in what follows, we will first 
discuss mass of asymptotically AdS spaces and then mass and 
angular momentum of asymptotically flat spaces.  
One can of course consider angular momentum in AdS backgrounds, but we 
will not do it here.

{\bf { Mass ( energy)  in AdS }}

Let us first show that for the time-like Killing vector 
$\bar{\xi}^\mu = ( -1, 0, 0, 0)$, which is related to the conserved energy, 
the terms involving dual Riemann tensors in (\ref{linearcotton}) 
do not contribute to $\sqrt{-\bar{g}} C^{0 0}\,_L \bar{\xi}_0$. We have
\begin{equation}
 \bar{\xi}_0\bar{\nabla}_\sigma v_\tau   
(\ast R^{\tau 0 \sigma \nu}\,_L +
\ast R^{\tau \nu \sigma 0}\,_L ) = 
-\bar{\xi}_0 \bar{\Gamma}^0\,_{\sigma \tau}v_0
\epsilon^{\sigma 0 \alpha \beta} \bar{g}^{\rho 0}  (R^\tau\,_{\rho \alpha \beta})_L
\label{dualpart}
\end{equation}
In the commonly used coordinates, the background AdS metric
\begin{equation}
ds^2= - ( 1- \frac{\Lambda}{3} r^2) dt^2 + {(1- \frac{\Lambda}{3} r^2)}^{-1} dr^2 + r^2 d\Omega_2 \, ,
\label{adsmetric}
\end{equation}
has non-vanishing  $\bar{\Gamma}^0\,_{0 r}$ and its symmetric partner. 
Given this and the fact 
that linearized Riemann tensor obeys the algebraic symmetries of the full Riemann tensor, we see 
that (\ref{dualpart}) does not contribute to the energy in AdS or flat backgrounds. 
on the other hand, the remaining non-trivial part can be expressed as ( let us keep  $\bar{\xi}_\nu$ for now and
choose it to be time-like at the very end )  
\begin{eqnarray}
-2 \sqrt{-\bar{g}} C^{\mu \nu}\,_L \bar{\xi}_\nu &=& \bar{\nabla}_\alpha \left(\, 
v_\sigma \bar{\xi}^\rho \epsilon^{\sigma \mu \alpha \beta} {\cal G}^L\,_{\rho \beta} + 
v_\sigma \bar{\xi}_\nu \epsilon^{\sigma \nu \alpha \beta} {\cal G}^L\,^\mu\,_\beta + 
v_\sigma \bar{\xi}_\nu \epsilon^{\sigma \mu \nu \beta} {\cal G}^L\,^\alpha\,_\beta \right) \nonumber \\
&&-  v_\sigma \epsilon^{\sigma \nu \alpha \beta} {\cal G}^L\,^\mu\,_\beta \bar{\nabla}_\alpha \bar{\xi}_\nu 
- v_\sigma \epsilon^{\sigma \mu \alpha \beta} {\cal G}^L\,_{\rho \beta}\bar{\nabla}_\alpha \bar{\xi}^\rho
  -  v_\sigma \epsilon^{\sigma \mu \nu \beta} {\cal G}^L\,^\alpha\,_\beta \bar{\nabla}_\alpha \bar{\xi}_\nu \, .
\nonumber
  \end{eqnarray}
The last two terms cancel each other due to the fact that $\bar{\xi}^\mu$ is a Killing vector. Hence one is 
left with
\begin{eqnarray} 
 -2 \sqrt{-\bar{g}} C^{\mu \nu}\,_L \bar{\xi}_\nu &=&  \bar{\nabla}_\alpha \left(\, 
v_\sigma \bar{\xi}^\rho \epsilon^{\sigma \mu \alpha \beta} {\cal G}^L\,_{\rho \beta} + 
v_\sigma \bar{\xi}_\nu \epsilon^{\sigma \nu \alpha \beta} {\cal G}^L\,^\mu\,_\beta + 
v_\sigma \bar{\xi}_\nu \epsilon^{\sigma \mu \nu \beta} {\cal G}^L\,^\alpha\,_\beta \right) \nonumber \\
&&- v_\sigma \epsilon^{\sigma \nu \alpha \beta} {\cal G}^L\,^\mu\,_\beta \bar{\nabla}_\alpha \bar{\xi}_\nu 
\label{boundarycot2}
\end{eqnarray} 
The first line, with an anti-symmetric tensor density inside the covariant derivative, is in the desired 
boundary form. The final piece can be handled in the following way: define 
\begin{equation} 
\bar{\Xi}^\mu \equiv  \frac{1}{\sqrt{-g}}\,v_\sigma \,\epsilon^{\sigma \nu \alpha \mu}
\bar{\nabla}_\alpha \bar{\xi}_\nu
\end{equation}
then the last term in (\ref{boundarycot2}) becomes
\begin{equation}
-\sqrt{-\bar{g}}{\cal G}^L\,^\mu\,_\beta \bar{\Xi}_\beta 
\end{equation}
which resembles the conserved charges in the pure cosmological Einstein theory except we still need to show that 
$\bar{\Xi}_\beta$ is a Killing vector. In fact,
\begin{equation}
\bar{\nabla}_\alpha \bar{\Xi}_\beta +  \bar{\nabla}_\beta \bar{\Xi}_\alpha = 0
\end{equation}  
follows once we employ a basic identity $\bar{\nabla}_\alpha \bar{\nabla}_\beta \bar{\Xi}_\nu = 
\bar{R}^\mu\,_{\nu \beta \alpha}\bar{\Xi}_\mu$. Thus the last term in 
(\ref{boundarycot2}) can also be written as a boundary term as in cosmological Einstein 
theory); we just need to replace $\bar{\xi}_\nu$ with $\bar{\Xi}_\nu$ in (\ref{einsteinboundary1}).
 
Finally let us summarize the mass formula in Chern-Simons Modified GR for asymptotically AdS ( or in fact asymptotically 
flat spaces as well.) It is more convenient to keep a covariant looking expression [One can easily  
choose $\xi^\nu= (-1,0,0,0)$.]
\begin{equation}
E = \frac{1}{16 \pi G} \, \oint_{S^2} \,
dS_{i} \, \left( {\cal{Q}}^{0 i}_{E} (\bar{\xi}) + \frac{1}{2} \, 
{\cal {Q}}^{0 i}_{E} (\bar{\Xi}) - \frac{1}{2} \, {\cal {Q}}^{0 i}_{C} (\bar{\xi})\right) \, ,
\label{csgrmass}
\end{equation}
where
\begin{eqnarray}
{\cal{Q}}^{\mu i}_{E} (\bar{\xi}) & \equiv & \sqrt{-\bar{g}} \left( 
\bar{\xi}_{\nu} \, \bar{\nabla}^{\mu} \, h^{i \nu} -
\bar{\xi}_{\nu} \, \bar{\nabla}^{i} \, h^{\mu\nu} +
\bar{\xi}^{\mu} \, \bar{\nabla}^{i} \, h -
\bar{\xi}^{i} \, \bar{\nabla}^{\mu} \, h \right.  \nonumber \\
& & \quad \qquad \left. + h^{\mu\nu} \, \bar{\nabla}^{i} \, \bar{\xi}_{\nu}
- h^{i \nu} \, \bar{\nabla}^{\mu} \, \bar{\xi}_{\nu}
+ \bar{\xi}^{i} \, \bar{\nabla}_{\nu} \, h^{\mu\nu}
- \bar{\xi}^{\mu} \, \bar{\nabla}_{\nu} \, h^{i \nu}
+ h \, \bar{\nabla}^{\mu} \, \bar{\xi}^{i} \right) \, , \label{einchar} \\
{\cal{Q}}^{\mu i}_{C} (\bar{\xi}) & \equiv & v_\sigma \bar{\xi}^\rho \epsilon^{\sigma \mu i \beta} {\cal G}^L\,_{\rho \beta} + 
v_\sigma \bar{\xi}_\nu \epsilon^{\sigma \nu i \beta} {\cal G}^L\,^\mu\,_\beta + 
v_\sigma \bar{\xi}_\nu \epsilon^{\sigma \mu \nu \beta} {\cal G}^L\,^i\,_\beta 
\end{eqnarray} [ The integration should be carried over a sphere at spatial infinity. So, de Sitter (as opposed to AdS), with a cosmological horizon, 
is ruled out \cite{ad,dt1}.] This result resembles to what we have found 
for the three dimensional TMG \cite{dt2}, but of course, now we have an 
additional background vector. 

It is important to check the background gauge invariance of our definition, 
which in our case follows easily. Consider small diffeomophisms generated by 
$\zeta_\mu$, such that 
\begin{equation}
\delta_\zeta h_{\mu \nu}= \bar{\nabla}_\mu \zeta_\nu + 
\bar{\nabla}_\nu \zeta_\mu \, .
\end{equation} 
One can show that (see the second paper in {\cite{dt1}}, though there is an important typo in that 
discussion ), 
${\cal {G}}^L\,_{\mu \nu}$ is (gauge) invariant under these transformations. 
Therefore, Einstein and Cotton parts are separately gauge invariant in our 
energy expression. 

Now that we have the mass formula in our hand we can find the mass of the known solutions. Unfortunately, in this model, the only 
solution we have is the AdS-Schwarzschild black hole, whose metric reads
\begin{equation}
ds^2= - ( 1- \frac{r_0}{r}- \frac{\Lambda}{3} r^2) dt^2 + {(1-  \frac{r_0}{r}- \frac{\Lambda}{3} r^2)}^{-1} dr^2 + 
r^2 d\Omega_2\, .
\label{adsblackholemetric}
\end{equation}
Explicit computation of the mass of this metric using our formula (\ref{csgrmass}) is actually straightforward: $\bar{\Xi}^\mu$ 
vanishes so ${\cal{Q}}_E(\bar{\Xi})$ term does not contribute. on the other hand, ${\cal{Q}}_C(\bar{\xi})$ term vanishes 
for various reasons ( such as symmetry and   ${\cal G}^L\,^\mu\,_\beta$ being zero for this Einstein space at infinity). 
From the first part, we get  $E = M$, if we identify $ r_0 = 2 G M$ as usual. This result is consistent and expected, as the metric 
received no correction from the Cotton part. But, if a non-trivial solution to the full CS modified GR is found, our formula 
will give its mass. In the three dimensional theory, for rotating solutions in AdS, one can see how  the Cotton term modifies 
the mass and angular momentum \cite{dkt,ost}. 
 
{\bf { Angular Momentum and energy in flat space }}

In flat space ( and in Cartesian coordinates), our task is somewhat simple: we can give a unified formula for both mass and 
angular momentum. All the covariant derivatives become ordinary derivatives. Without going into further details let us 
write the final answer
\begin{equation}
Q^0(\bar{\xi}_\mu) = \frac{1}{16 \pi G} \, \oint_{S^2} dS_i \Big( \bar{\xi}_0 (\partial_j h^{ij} - \partial^i h^j\,_j) + \bar{\xi}^i\partial_j h^{0j}
- \bar{\xi}_j\partial^i h^{0 j} +\frac{\mu}{2}\epsilon^{i j k}\bar{\xi}_j{\cal G}^L\,^0\,_k \Big)\, .
\end{equation}
For $\bar{\xi}_0 = ( 1, 0, 0,0)$,  $Q^0(\bar{\xi}_0)= E$ and 
and the formula reduces to the usual ADM one. In the case of the 
angular momentum, $\bar{\xi}_i = ( 0, 0, 0, 1 )$ and $Q^0(\bar{\xi}_i)= J$: 
There {\it{is}} a contribution from the Cotton part. 
Currently, since we do not know any rotating  solution in this model, 
we cannot give an explicit example.

\section{\label{conc} Conclusions}

We have constructed the conserved energy of the 
Chern-Simons modified General Relativity \cite{jp} for asymptotically 
flat and asymptotically AdS spaces. We also provided the angular momentum 
for asymptotically flat spaces. Our construction follows \cite{ad} and 
\cite{dt1,dt2} and can be generalized to the models with additional higher 
curvature terms. The theory we have considered has an external field which 
depends on time and breaks some part of the Lorentz group. But constant 
time translations and spatial rotations are still symmetries of the model 
which allowed us to define the total mass and angular momentum of 
the spacetimes that solve the field equations. Whether or not CS modified 
GR is physical or not is not yet clear since the novel predictions of the 
theory have been related to weak gravity regime where the current  
experiments are of no help. Two such examples are : The theory predicts 
parity-violating gravitational waves \cite{jp} and parity-violating 
interaction between gravitoelectric and gravito-magnetic fields \cite{seck}

\section{\label{ackno} Acknowledgments}

I would like to thank S. Deser, \" O. Sar{\i}o\u{g}lu and  A. Waldron 
for useful discussions and the Mathematics Department 
at the University of California   
Davis for hospitality. This work was supported by the ``Young 
Investigator Fellowship'' of the Turkish Academy of Sciences (T\"{U}BA) 
and by the  T\"{U}B\.{I}TAK Kariyer Grant No 104T177.


\newpage

\begin{thebibliography}{99}
\bibitem{jp}  R. Jackiw and S. Y. Pi, Phys. Rev.  D {\bf 68}, 104012 
(2003).
\bibitem{djt} S. Deser, R. Jackiw and S. Templeton, Phys. Rev. Lett. {\bf 48}
975 (1982); Ann. Phys. (N.Y.) {\bf 140} 372 (1982); {\bf 185} 406(E) (1988).
\bibitem{cfj} S.~M.~Carroll, G.~B.~Field and R.~Jackiw, Phys. Rev.  
D {\bf 41}, 1231 (1990).
\bibitem{adm} R.~Arnowitt, S.~Deser and C.~W.~Misner, Phys. Rev {\bf 116}, 
1322 (1959); {\bf 117}, 1595 (1960); ``The dynamics of general relativity,''
arXiv:gr-qc/0405109.
\bibitem{ad}L.~F.~Abbott and S.~Deser, Nucl. Phys. B {\bf 195}, 76 (1982).
\bibitem{dt1} S. Deser and B. Tekin, Phys. Rev. Lett. {\bf 89} 101101 (2002);
Phys. Rev. D {\bf 67} 084009 (2003). 
\bibitem{dt2} S. Deser and B. Tekin, Class. Quantum Grav. {\bf 20} L259 
(2003).
\bibitem{dkt} S. Deser, \.{I}. Kan{\i}k and B. Tekin, Class. Quantum Grav. 
{\bf 22} 3383 (2005).
\bibitem{gibbons} G.~W.~Gibbons, H.~Lu, D.~N.~Page and C.~N.~Pope,
  J.\ Geom.\ Phys.\  {\bf 53}, 49 (2005). 
\bibitem{ost} S.~Olmez, O.~Sarioglu and B.~Tekin,
  Class.\ Quant.\ Grav.\  {\bf 22}, 4355 (2005).
\bibitem{hete} M.~Henneaux and C.~Teitelboim,
  Commun.\ Math.\ Phys.\  {\bf 98}, 391 (1985).
\bibitem{jackiw}
  R.~Jackiw, ``Lorentz Violation in a Diffeomorphism-Invariant Theory,''
  arXiv:0709.2348 [hep-th].
\bibitem{gh} D.~Guarrera and A.~J.~Hariton,
  Phys.\ Rev.\  D {\bf 76}, 044011 (2007).  
\bibitem{mariz1}T.~Mariz, J.~R.~Nascimento, E.~Passos and R.~F.~Ribeiro,
  Phys.\ Rev.\  D {\bf 70}, 024014 (2004).
\bibitem{mariz2} T.~Mariz, J.~R.~Nascimento, A.~Y.~Petrov, L.~Y.~Santos 
and  A.~J.~da Silva,
``Lorentz violation and the proper-time method,''
  arXiv:0708.3348 [hep-th].
\bibitem{cdko}
  B.~A.~Campbell, M.~J.~Duncan, N.~Kaloper and K.~A.~Olive,
  Nucl.\ Phys.\  B {\bf 351}, 778 (1991).
\bibitem{seck} T.~L.~Smith, A.~L.~Erickcek, R.~R.~Caldwell and M.~Kamionkowski,
  ``The effects of Chern-Simons gravity on bodies orbiting the Earth,''
  arXiv:0708.0001 [astro-ph].
\bibitem{kmt}
  K.~Konno, T.~Matsuyama and S.~Tanda,
  Phys.\ Rev.\  D {\bf 76}, 024009 (2007).

\end{thebibliography}
\end{document}